\documentclass[11pt,a4paper]{article}

\usepackage{amsmath}
\usepackage{amssymb}
\usepackage{setspace}
\usepackage{subfigure}
\usepackage[usenames,dvipsnames]{color} 
\usepackage{marvosym}
\usepackage[usenames]{color}
\usepackage[pdftex]{graphicx}
\usepackage{epstopdf}
\title{The First Passage Probability of Intracellular Particle Trafficking}
\author{Salman S. Rogers, Neftali Flores-Rodriguez, \\Victoria J. Allan, Philip G. Woodman, Thomas A. Waigh
\thanks{thomas.waigh@manchester.ac.uk}\\
University of Manchester}

\begin{document}

\maketitle
\abstract{
The first passage probability (FPP), of trafficked intracellular particles reaching a displacement $L$, in a given time $t$ or inverse velocity $S=t/L$, can be calculated robustly from measured particle tracks, and gives a measure of particle movement in which different types of motion, e.g.~diffusion, ballistic motion, and transient run-rest motion, can readily be distinguished in a single graph, and compared with mathematical models. The FPP is attractive in that it offers a means of reducing the data in the measured tracks, without making assumptions about the mechanism of motion: for example, it does not employ smoothing, segementation or arbitrary thresholds to discriminate between different types of motion in a particle track. Taking experimental data from tracked endocytic vesicles, and calculating the FPP, we see how three molecular treatments affect the trafficking. We show the FPP can quantify complicated movement which is neither completely random nor completely deterministic, making it highly applicable to trafficked particles in cell biology.
}

\section{Introduction}
The first passage probability $F(t,L)$, is the probability density per unit time, that a moving particle will require a time $t$ before it exceeds a displacement $L$, \textit{for the first time}. The concept of FPP has a long history in the study of stochastic processes such as Brownian motion and chemical reaction kinetics \cite{redner2001}. However, we report here its utility in analysing the trafficking of particles in the cell, whose motion is neither random nor deterministic, but complicated by many partially understood effects.

The FPP can be calculated for both observed particle tracks and theoretical models of particle motion, for a range of values of the parameter $L$, and potentially offers a way of analysing the particle motion, to characterise relevant velocities, diffusivities, or length and time scales. Some advantages of using the FPP are that it is a robust measure of particle motion, and avoids the use of smoothing and segmentation of particle tracks, or the use of thresholds to discriminate between the different kinds of motion, such as runs and rests. It therefore avoids the artifacts that smoothing and thresholding produce. For instance we may be interested in the trafficking of a set of fluorescently-labelled particles which are driven intermittantly by molecular motors on microtubules and F-actin, but these may also experience advection in the flowing cytoplasm, and Brownian motion. A simple way of analysing such complicated motion is to reduce its noise by spatial and temporal filtering, and break it up into different components \cite{arcizet2008,bacher2004,debruin2007,huet2006,Miura2005,pangarkar2005}. These procedures are useful and conceptually straight-forward, but have obvious disadvantages in that they all involve ``fudge factors'', such as smoothing length scales and time scales, or thresholds to distinguish how fast, straight or long-lived a track must be in order to count as a segment of motion we want to measure. These procedures weaken the analysis since the measured tracks may be far more complicated than the ideals we have in mind, and it is often unclear how much the final result depends on the arbitrary parameters and implicit assumptions.

The present study is motivated by experimental data from fluorescently-labeled Rab5 GTPase, which plays a role in the coordination of vesicle trafficking in the endocytic pathway of eukaryotic cells \cite{nielsen1999ncb}. Having previously developed a method of tracking hundreds of vesicles moving simultaneously in the cell \cite{rogers2007}, we began to explore methods of analysing their motion. The tracks are characterised by several qualitative features: vesicles make directed runs at a range of velocities; these runs persist for variable times and are interupted by rests of variable times; after a rest, the vesicles sometimes carry on with motion in the same direction, but often make abrupt changes of direction, or seem to reverse their motion along a linear track; and runs are sometimes straight or gently curved, but often follow a more erratic path.

In this article, we demonstrate the calculation of the FPP from experimental data taken from control cells as well as cells treated with nozodazole, which disrupts the microtubules, thus compromising the motility of Rab5 GTPase. We compare the results with simple theoretical models of intracellular trafficking, and a simple segmentation analysis of the tracks. In a separate article, now in preparation, we will present new results on the intracellular trafficking of vesicles, using the first passage probability distribution, as well as other analyses.

\section{Experimental section}
\subsection{Cell culture and live cell imaging}
HeLaM cells, in DMEM containing 10\% FBS, were grown in 35mm glass bottomed dishes (MatTek Corporation, Ashland, MA, USA) and imaged at 37$^\circ$C on an Olympus IX81 microscope using a 100x objective lens, 1.35 NA and an additional 1.5x lens magnifier, fitted with a Optoscan high speed dynamic bandpass control monochromator (1800 g mm$^{-1}$ Holographic grating) (Cairn Research, Faversham, UK), and a Photometrics Cascade 512 back-illuminated camera (Photometrics, Tuscon, AZ, USA). Cells were imaged for 500 frames using continuous illumination at 10 frames/s. Control experiments showed that imaging under these conditions for as long as 50 minutes did not affect the probability that the cells would successfully divide, or undergo apoptosis, in the following 24 hours (data not shown).

Cells were treated with nocodazole, to depolymerise microtubules, as follows. A dish of HeLaM cells were incubated in DMEM containing 1-$\mu$M nocodazole, at 4$^\circ$C for 5 minutes, followed by 2 hours at 37$^\circ$C. The cells were then imaged as above.

\subsection{Particle tracking}
Our recently developed tracking method, the polynomial-fit, Gaussian-weight algorithm (PFGW) \cite{rogers2007}, allows us to accurately track the extrema of intensity corresponding to individual vesicles, without errors due to the presence of their neighbours in the image. Thresholds of acceptable eccentricity, radius, skewness and particle lifetime were employed to reject intensity extrema that did not correspond to single vesicles. Manual examination of the tracked movies showed that all particles were tracked except for a small minority of faint fast-moving particles. Fig.~\ref{seg}(a) shows example tracks for the control cells.

\begin{figure}
\centering
\resizebox{8cm}{!}{\includegraphics{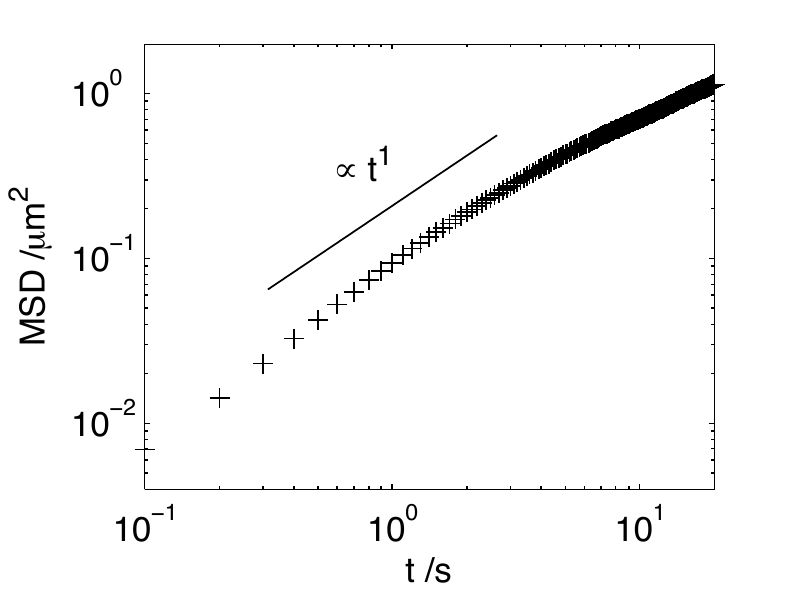}}
\caption{Mean square displacement as a function of time, MSD($t$), for all vesicle tracks in the control cells. The static error is negligible, and the average motion is superdiffusive at $t\lesssim1$~s, and subdiffusive at $t\gtrsim1$~s.}
\label{MSD}
\end{figure}

Static errors may be estimated by calculating the mean square displacement (MSD), as a function of time scale $t$, for the measured tracks \cite{rogers2007,savin2005}. Fig.~\ref{MSD} shows the MSD for the control cells. At the shortest time scale of 0.1~s, there is no plateau that would be characteristic of static error. Therefore, the average static error in measuring displacements is significantly less than $\sqrt{\textrm{MSD}(0.1~\textrm{s})}=0.08~\mu$m. It is also apparent from Fig.~\ref{MSD} that the MSD has an exponent greater than 1 at $t\lesssim 1$~s and less than 1 at $t\gtrsim1$~s. Thus the vesicle motion is superdiffusive at short time scales---i.e.~dominated by active motion, and subdiffusive at long time scales---presumably because the vesicles tend to stop moving or lose their initial direction after times of the order of $\sim 1$~s.

\subsection{Segmentation analysis}
\begin{figure}
\centering
\subfigure[]{\resizebox{8cm}{!}{\includegraphics{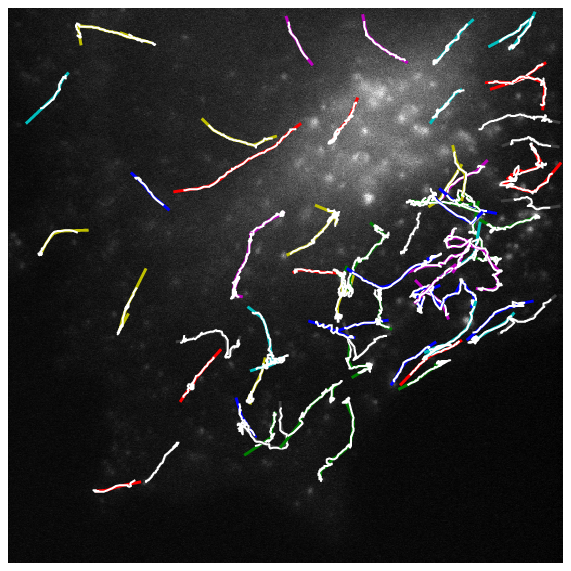}}}
\subfigure[]{\resizebox{8cm}{!}{\includegraphics{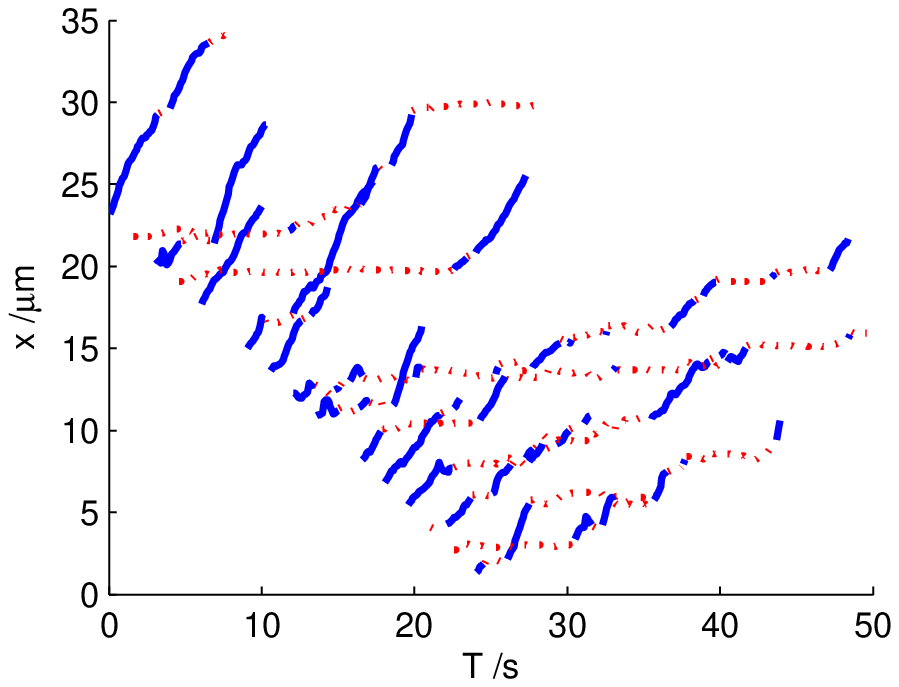}}}
\caption{(a) Example tracks (white) and smoothed contours (coloured online) of all tracks of total contour length greater than 3~$\mu$m, in one video of the control cells. (b) Examples of parametrised distance along the smoothed contours $x(T)$ for the control cells, segmented into runs (solid, blue) and rests (dotted, red).}
\label{seg}
\end{figure}

Apart from the first passage probability calculation, we analysed the tracks by a method of segmenting each particle track into discrete runs and rests. This heuristic method is similar to methods used to characterise intracellular motion by several previous authors. We briefly detail our particular segmentation analysis here. The analysis is based on the observation that the vesicles tend to make directed \emph{runs} along straight or gently curved trajectories, which are punctuated by stationary \emph{rests}. The length, duration and average speed of each run or rest is measured by: 

\begin{enumerate}
\item A smoothed contour is created to correspond to each particle track. The first step was to replace the coordinates of each point in the particle track with the mean of all coordinates in the track which were within a threshold distance of $L_\textit{pix}=4$ to that point. Thus the track is smoothed spatially rather than temporally: this spatial averaging has the advantage that short runs of directed motion within a track are more preserved than if we took a running average in time of points along the track. The second step was to produce a coarse version of the track by successively taking sets of points in the smoothed track separated by the distance $L_\textit{pix}$, starting from the initial point, and replacing each set with its mean position. Finally, the initial and final displacements in the coarsened track were extrapolated so that they extend further than the initial and final points of the original track. Examples of track contours can be seen in Fig.~\ref{seg}(a): they consist of straight-line sections of length $\sim L_\textit{pix}$.
\item The position $x(T)$ of a particle along its contour, was calculated as the projection of the smoothed track along the contour, taking the projection of its initial coordinates as $x=0$. 
\item The parametrised distance $x(T)$ is segmented into discrete runs and rests. Wherever the particle moves less than 1 pixel in more than 5 frames, those positions are marked as a rest. The remaining sections are counted as runs if the total displacement of each is more than 1 pixel. Fig.~\ref{seg}(b) shows examples of $x(T)$ for the control data, segmented into solid (blue) sections for runs and dotted (red) for rests.
\end{enumerate}
This procedure ignores other effects observed in the motion of the particles, such as reversals of direction and slower erratic motion, and has advantages and disadvantages as mentioned below and in our forthcoming article \cite{inpreparation2009}.

\section{Results}
Rab5-labelled vesicles were tracked in control cells, as well as cells subjected to nocodazole treatment, as described above. For each case of control or treated cells, at least 10 videos, of 500 frames each, were captured of separate cells. The average numbers of particles simultaneously tracked at any instant in the control and nocodazole treatment, were 128 and 54 respectively.

From a set of observed tracks $\{\mathbf{R}_n(T)\}$, the first passage probability distribution is calculated by finding the smallest non-negative $t$ that satisfies $|\mathbf{R}_n(T+t)-\mathbf{R}_n(T)|=L$, at each starting time point $T$, at each track in the set, indexed by $n$. To calculate $F(t,L)$, these values of $t$ are counted in bins; then the count is normalised and plotted as a histogram. 

The first passage probability can also be calculated in terms of an inverse velocity: $F(S,L)$, where $S=t/L$ is the inverse velocity or ``slowness''. Since $F(S,L) ds = F(t,L) dt$, the distributions can be converted by scaling $t$ with the length scale: 

\begin{equation}
F(S,L) = LF(t,L)|_{t=SL} \,. 
\label{implicitTransformation}
\end{equation}

\begin{figure}
\centering
\resizebox{16cm}{!}{\includegraphics{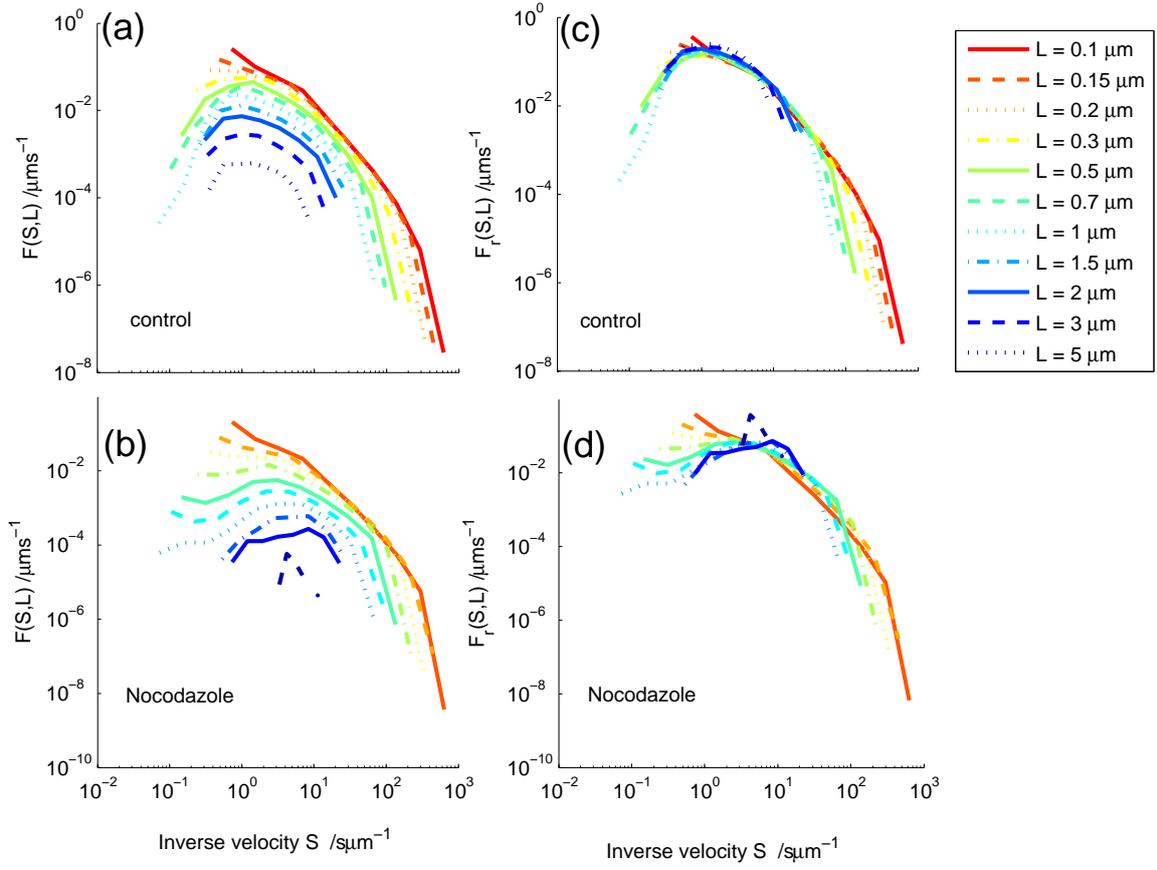}}
\caption{The first passage probability distribution for all data sets, at a range of $L$. The distributions are normalised (a--b) by the number of data points in all tracks, and (c--d) renormalised by the number of passages at each $L$. The distributions for all treatments show a single peak which becomes narrower as $L$ increases.}
\label{fpp_all}
\end{figure}

Thus to calculate $F(S,L)$, the set of first passage times $t$ is replaced with $S=t/L$, before binning and counting. Fig.~\ref{fpp_all} shows $F(S,L)$ for each cell treatment. The distribution is normalised by the total number of recorded time points: thus the area under each curve is the total probability that a vesicle will make a passage of length $L$ from any starting point. It is useful to renormalise this distribution to form a new distribution, $F_r(S,L)$, in which the area under each curve is unity. The difference between $F$ and $F_r$ reflects the fact that the tracks are finite and often much shorter than the length scales in which we are interested: the renormalised distributions only count starting points that result in a passage. The graphs are qualitatively similar in that they show singly peaked distributions, with maxima in the range $S$=1--10~s~$\mu$m$^{-1}$. As $L$ increases, the distributions become narrower, especially in the case of the nocodazole treatment. 

\begin{figure}
\centering
\resizebox{16cm}{!}{\includegraphics{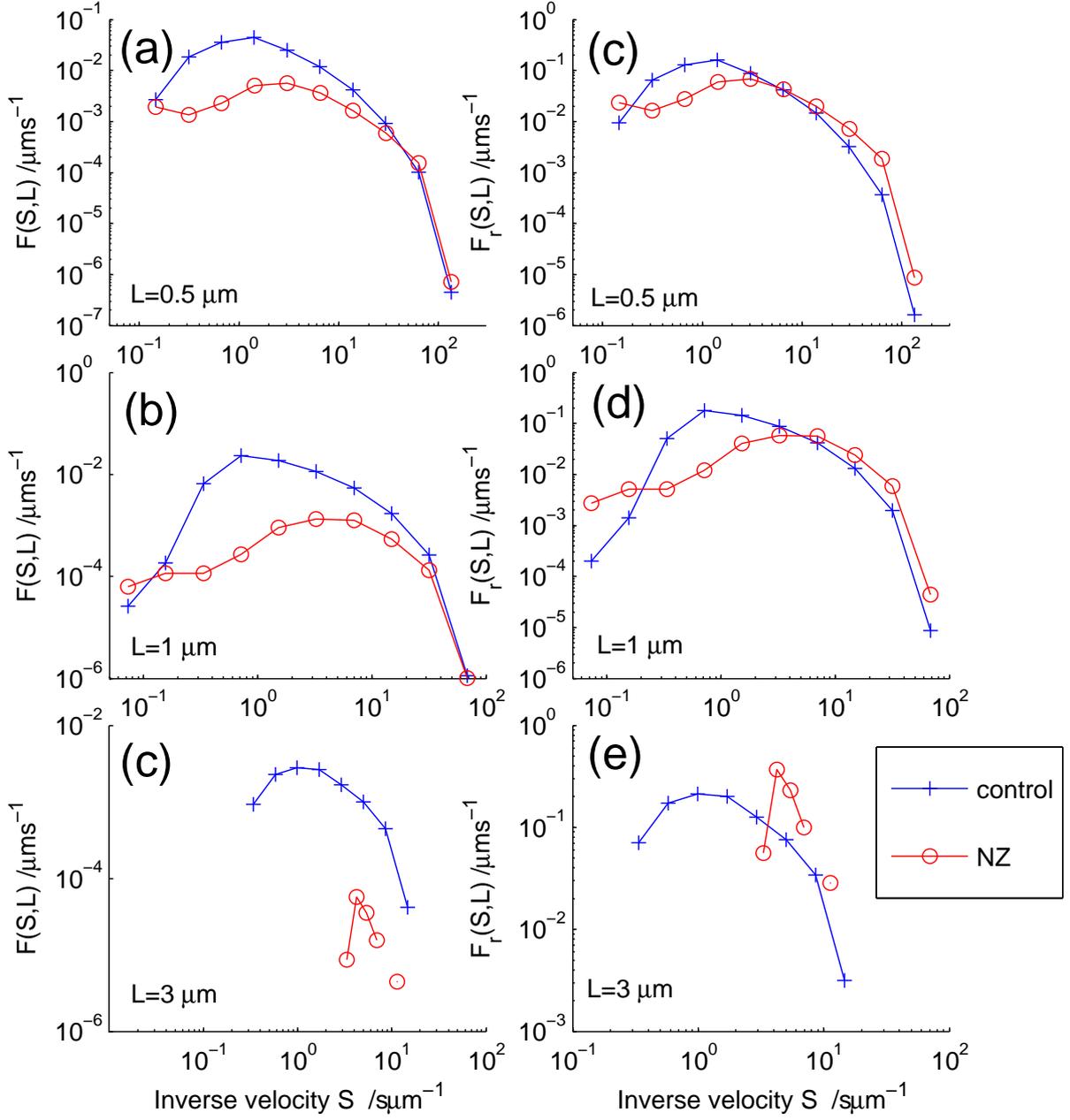}}
\caption{The first passage probability distribution for all data sets, at $L=0.5$, 1, 3~$\mu$m. Both normalisations are shown: $F(S,L)$ and $F_r(S,L)$ (a--c and d--f). Probabilities are markedly reduced under the nocodazole treatment compared to the control cells.}
\label{compare_fpp}
\end{figure}

We can also plot the first passage probabilities to compare the different treatments. Fig.~\ref{compare_fpp} shows $F(S,L)$ at $L=$0.5, 1 and 3~$\mu$m respectively. As the length scale increases, the difference between the treatments becomes more apparent. For example, if we consider an inverse velocity of $S=1$~s~$\mu$m$^{-1}$, corresponding to the peak FPP of the control cells, the probability of vesicles traversing various distances, at the corresponding speed of $1/S=1~\mu$m~s$^{-1}$, is markedly reduced under the nocodazole treatment compared to the control cells (Table~\ref{ref_fpp}). The differences between values of $F(1~$s~$\mu$m$^{-1},L)$ are greater than between $F_r(1~$s~$\mu$m$^{-1},L)$, since the proportion of vesicles making passages of 1--3~$\mu$m is significantly reduced by the nocodazole treatment. $F(1~$s~$\mu$m$^{-1},1~\mu$m$)$ is a factor of 10--50 lower for the treated cells than the control cells. In all graphs of Fig.~\ref{compare_fpp}, we can see that the nocodazole treatment impairs motion at lower $S$---i.e.~the probability of particles moving at larger velocities is markedly reduced. The difference increases with increasing $L$: i.e.~motion is increasingly impaired at larger length scales.

\begin{table}
  \begin{tabular}{l | c  c  c | c  c  c }
      & \multicolumn{3}{|l}{$F(1~$s$~\mu$m$^{-1},L)$} & \multicolumn{3}{|l}{$F_r(1~$s$~\mu$m$^{-1},L)$} \\
	\cline{2-4} \cline{5-7}
    L/$\mu$m & 0.5 & 1 & 3 & 0.5 & 1 & 3 \\
    \hline
	Control & $4.0\times 10^{-2}$ & $2.1\times 10^{-2}$ & $2.8\times 10^{-3}$ & $1.4\times 10^{-1}$ & $1.6\times 10^{-1}$ & $2.1\times 10^{-1}$ \\
    Nocodazole & $3.5\times 10^{-3}$ & $4.6\times 10^{-4}$ & $0$ & $4.2\times 10^{-2}$ & $2.1\times 10^{-2}$ & $0$ \\
    
  \end{tabular}
\caption{Reference values of first passage probability for all cell treatments.}
\label{ref_fpp}
\end{table}

\begin{figure}
\centering
\resizebox{!}{6cm}{\includegraphics{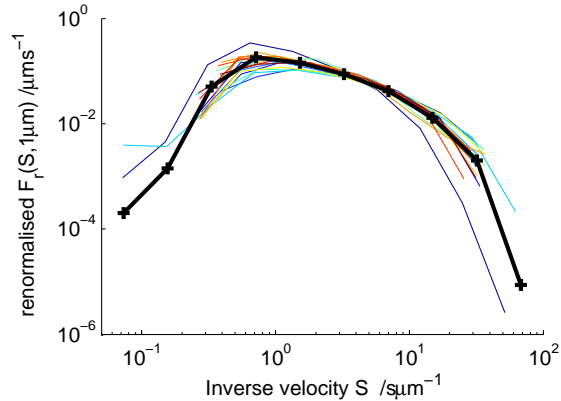}}
\caption{$F(S,1~\mu$m$)$ for the control cells: from each video (thin lines, coloured online) and from all videos combined (solid line ---$\mkern-15mu$\small{+}~).}
\label{variability}
\end{figure}

In order to test the reproducibility of behaviour between cells, we plot $F(S,1~\mu$m$)$ for the tracks from each video from the control data, as well as the combined distribution in Fig.~\ref{variability}. The differences in FPP between the control cells are due to the variability in behaviour of cells as well as the statistical error of sampling the vesicle tracks. However the differences are much smaller than those caused by the nocodazole treatment as shown above.

\section{Theoretical models}
The observed first passage probability distributions can be compared with theoretical models of various types of particle motion. We consider three simple models below, from which analytical expressions for the FPP can be calculated, then expand the discussion to more complicated types of motion, in comparison with the experimental data.

\subsection{Steady directional motion}
We consider first the simplest possible kind of particle motion. If all particles were to move at a constant speed $v$, with each moving in any constant direction, then the first passage probability would be a delta function, since the time for a particle to move $L$ would always be $L/v$:
\begin{equation}
F(S,L)=\delta (S-1/v) \,.
\end{equation}
Similarly, if each particle moved at a constant speed in any direction, where its speed was chosen from a probability distribution $P(v)$, then
\begin{equation}
F(S,L)=P(v) \left|\frac{dv}{dS}\right| = \frac 1 {S^2} P(1/S)\,.
\end{equation}
In either case, the first passage probability, as a function of $S$, is independent of $L$. This does not match $F(S,L)$ from the observed data which becomes narrower with increasing $L$. Indeed steady directional motion obviously does not match the captured videos, in which the vesicles show saltatory and erratic motion.

\subsection{Random walks}
\begin{figure}
\centering
\resizebox{8cm}{!}{\includegraphics{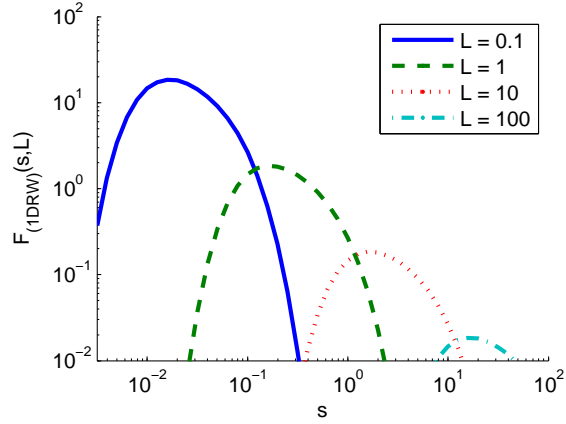}}
\caption{The first passage probability for the one-dimensional random walk, as a function of inverse speed, at a range of $L$, taking $D=1$. The peak position scales linearly with $L$.}
\label{randomWalkModel}
\end{figure}

For a random walk characterised by a diffusivity $D$, the probability $C(t,\mathbf{r})$ of finding the particle at $\mathbf{r}$ at time $t$ is governed by the diffusion equation:
\begin{equation}
\frac{\partial C}{\partial t}=D\nabla^2 C \,.
\label{eq:diff}
\end{equation}
If the particle is at the origin at $t=0$, i.e.~$C(0,\mathbf{r})=\delta(\mathbf{r})$, then Eq.~\ref{eq:diff} can be solved to yield the occupational probability $C_\textrm{(O)}(t,\mathbf{r})$ at subsequent times \cite{redner2001}:
\begin{equation}
C_\textrm{(O)}(t,\mathbf{r})=\frac{e^{-\mathbf{r}^2/4Dt}}{(4\pi Dt)^{n/2}} \,,
\end{equation}
where $n$ is the dimensionality of the random walk. However, we must note that the first passage probability is different from the occupational probability: the latter includes the possibility that the particle, after it has reached a displacement $L$, could return to that $L$. The first passage probability can therefore be calculated from Eq.~\ref{eq:diff}, using the same initial condition, but imposing a boundary condition $C(|\mathbf{R}|=L)=0$. This represents a sink of probability wherever the particle reaches a displacement of $L$; $C$ now includes only the trajectories that have not yet passed $L$. In one dimension, these considerations lead to the series solution for the probability of finding the particle before the first passage $C_\textrm{(BFP)}$  \cite{redner2001}:
\begin{equation}
C_\textrm{(BFP)}(t,x)=\frac{1}{L}\sum^{\infty}_{k=0}\cos\left(\frac{(2k+1)\pi x}{2L}\right) \exp \left[-\left(\frac{(2k+1)\pi}{2L}\right)^2 Dt\right]\,,
\end{equation}
The first passage probability is then the probability flux across the boundary:
\begin{equation}
F(t,L)=-D\oint_{|\mathbf{r}|=L} (\nabla C)\cdot d\mathbf{A} \,,
\end{equation}
or in one dimension:
\begin{equation}
F(t,L)=2D\left|\frac{\partial C}{\partial x}\right|_{x=L}\,.
\end{equation}
For the one dimensional random walk, the FPP is therefore:
\begin{equation}
F_{\textrm{(1DRW)}}(t,L)=\frac{\pi D}{L^2}\sum^{\infty}_{k=0} (-1)^k (2k+1) \exp \left[-\left(\frac{(2k+1)\pi}{2L}\right)^2 Dt\right] \,,
\end{equation}
or in terms of $S$:
\begin{equation}
F_{\textrm{(1DRW)}}(S,L)=\frac{\pi D}{L}\sum^{\infty}_{k=0} (-1)^k (2k+1) \exp \left[-\left(\frac{(2k+1)\pi}{2}\right)^2 DS/L\right] \,.
\label{eq:1drw}
\end{equation}
In Fig.~\ref{randomWalkModel}, we plot $F_{\textrm{(1DRW)}}(S,L)$ for a range of $L$, taking $D=1$. As shown, $F_{\textrm{(1DRW)}}(S,L)$ has a single peak which scales as $S\sim L$. The scaling is evident from Eq.~\ref{eq:1drw}, in which $F_{\textrm{(1DRW)}}(S,L)$ is a function of $DS/L$. For random walks in two or more dimensions, the first passage probability is qualitatively similar, and also scales as $S\sim L$.

By glancing at Figs.~\ref{randomWalkModel} and \ref{fpp_all}, we can see that the FPP from the measured data bears little resemblence to the random walk model, because the peak position in the measured data stays approximately constant although $L$ is increased by a factor of 50.

\subsection{Unidirectional runs and rests}
Suppose a particle alternately runs and rests. All runs share a constant direction and velocity $v$, and while running the particle has constant probability $a$, per unit time, of switching to a rest. During a rest, the particle does not move, but has constant probability $b$, per unit time of switching to another run. These single-particle switching rates $a$ and $b$ lead to a dynamic equilibrium where, over a long time, the particle will spend a proportion $f=(a/b+1)^{-1}$ of the time running, and $1-f=(b/a+1)^{-1}$ resting. 

The first passage probability distribution of this ``run-rest'' model, $F_{\textrm{(RR)}}(S,L)$, should have two limits which we can predict before we analyse the model in detail. Over a long track length $L$, the particle spends a total time $L/v$ running, thus the mean number of rests in this period is $\lambda=aL/v$. If $\lambda\ll 1$, then, by definition, the first passage probability is dominated by the runs: i.e.~$F_{\textrm{(RR)}}(S,L) \approx \delta(S-v^{-1})$ for $L\ll v/a$. However, if $\lambda\gg 1$, the particle experiences many rests and runs; as $\lambda\rightarrow \infty$, the proportion of time it spends running tends to $f$. Therefore $F_{\textrm{(RR)}}(S,L) \approx \delta(S-(vf)^{-1})$ for $L\gg v/a$. Thus we expect to see the first passage probability switching from a sharp peak at small-$L$ to another sharp peak at large-$L$.

To calculate $F_{\textrm{(RR)}}$ in detail, consider a passage of length $L$ that contains $n$ rests. Its total time will be
\begin{equation}
t=\frac Lv +\sum^{n}_{k} \tau_k \,,
\label{eq:runresttime}
\end{equation}
if each rest persists for a time $\tau_k$. Since there is a constant probability per unit time of the rest ending, the probability of that rest surviving for $\tau_k$ is distributed exponentially:
\begin{equation}
P_\textrm{rest}(\tau_k)=b e^{-b\tau_k}\,,
\end{equation}
and therefore, the probability of a particular combination of rest times $\{\tau_k\}$ that satisfy Eq.~\ref{eq:runresttime} is the compound probability
\begin{equation}
P(\{\tau_k\})=\prod^{n}_{k=1} b e^{-b\tau_k} = b^n e^{-b(t-L/v)}\,,
\label{eq:ptauk}
\end{equation}
making use of Eq.~\ref{eq:runresttime}. 

The probability of having $n$ rests in the length $L$ follows the Poisson distribution, since they occur independently of each other:
\begin{equation}
P(n)=\frac {e^{-\lambda} \lambda^n}{n!} \,,
\end{equation}
where $\lambda$ is given above. 

The probability that the passage requires a time $t$, given that it contains $n$ rests, is the integral of $P(\{\tau_k\})$ over all the possible combinations of $\{\tau_k\}$ that satisfy  Eq.~\ref{eq:runresttime}:
\begin{equation}
P(t|n)= \left\{ \begin{array}{ll} \delta(t-L/v) & ; n=0 \\
	            \int P(\{\tau_k\}) d^{n-1}\tau & ; n\ge 1 \,. \end{array} \right.
\end{equation}
This integral becomes an integral over a constant once we have substituted Eq.~\ref{eq:ptauk}:
\begin{equation}
P(t|n)= \frac{(t-L/v)^{n-1}}{(n-1)!} b^n e^{-b(t-L/v)} ~~ ; n\ge 1 \,.
\end{equation}
The probability that the passage requires a time $t$ is then obtained by a sum of $P(t|n)$ over all the possible values of $n$:
\begin{equation}
P(t)= \sum^{\infty}_{n=0} P(t|n) P(n) = e^{-\lambda} \delta(t-L/v) + \sum^{\infty}_{n=1} \frac{b^n(t-L/v)^{n-1} e^{-b(t-L/v)} e^{-\lambda} \lambda^n }{n!(n-1)!} \,.
\end{equation}
This series solution can be rearranged using the form of the first modified Bessel function of the first kind, $I_1$:
\begin{equation}
P(t)= e^{-aL/v} \delta(t-L/v) + \sqrt{\frac{abL/v}{t-L/v}} e^{-bt-(a-b)L/v}
	I_1\left( 2 \sqrt{ab\frac Lv \left(t-\frac Lv\right)}\right) \,.
\end{equation}
Finally, to obtain the first passage probability, we must consider the starting point of the particle, since it may be initially running or resting. In the latter case we must include the time of the initial rest:
\begin{align}
F_{\textrm{(RR)}}(t,L) &= fP(t) + (1-f) \int^{t-L/v}_{0} P_\textrm{rest}(\tau) P(t-\tau) d\tau \,, \\
			  &\begin{array}{ll} =& e^{-aL/v}\delta(t-L/v) + \frac{e^{-bt-(a-b)L/v}}{a/b+1}  \\
				&\times \left(aI_0(2\sqrt{ab(t-L/v)L/v})+\sqrt{\frac{abL/v}{t-L/v}} I_1(2\sqrt{ab(t-L/v)L/v})\right)\,,
				\end{array}
\end{align}
or in terms of $S$:
\begin{equation}
\begin{array}{ll} F_{\textrm{(RR)}}(S,L) =& Le^{-aLv^{-1}}\delta(S-v^{-1}) + \frac{e^{-bLs-(a-b)Lv^{-1}}}{a/b+1} \times  \\
			& \left(aI_0(2\sqrt{abL^2v^{-1}(S-v^{-1})})+\sqrt{\frac{abv^{-1}}{S-v^{-1}}} I_1(2\sqrt{abL^2v^{-1}(S-v^{-1})})\right)\,.
\end{array}
\label{eq:rr}
\end{equation}

\begin{figure}
\centering
\resizebox{15cm}{!}{\includegraphics{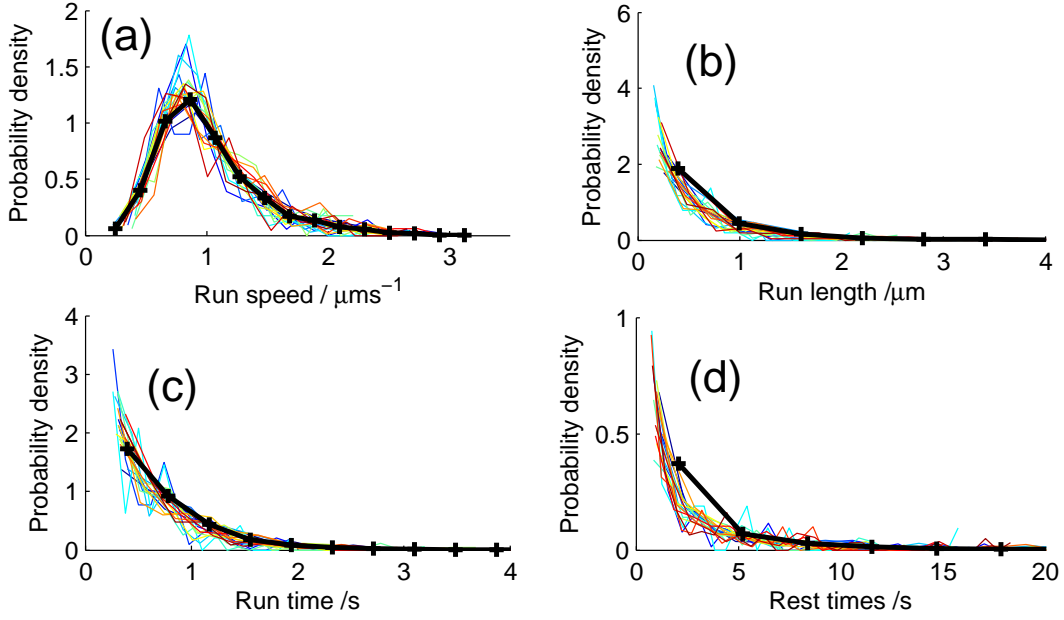}}
\caption{Statistics of segmented runs and rests from the control data: (a) run speed, (b) run length, (c) run time, (d) rest time. Distributions from of individual videos are shown (thin lines, coloured online) as well as the distributions of all videos combined (solid line ---$\mkern-15mu$\small{+}~).}
\label{segmentation}
\end{figure}

\begin{figure}
\centering
\subfigure[]{\resizebox{8cm}{!}{\includegraphics{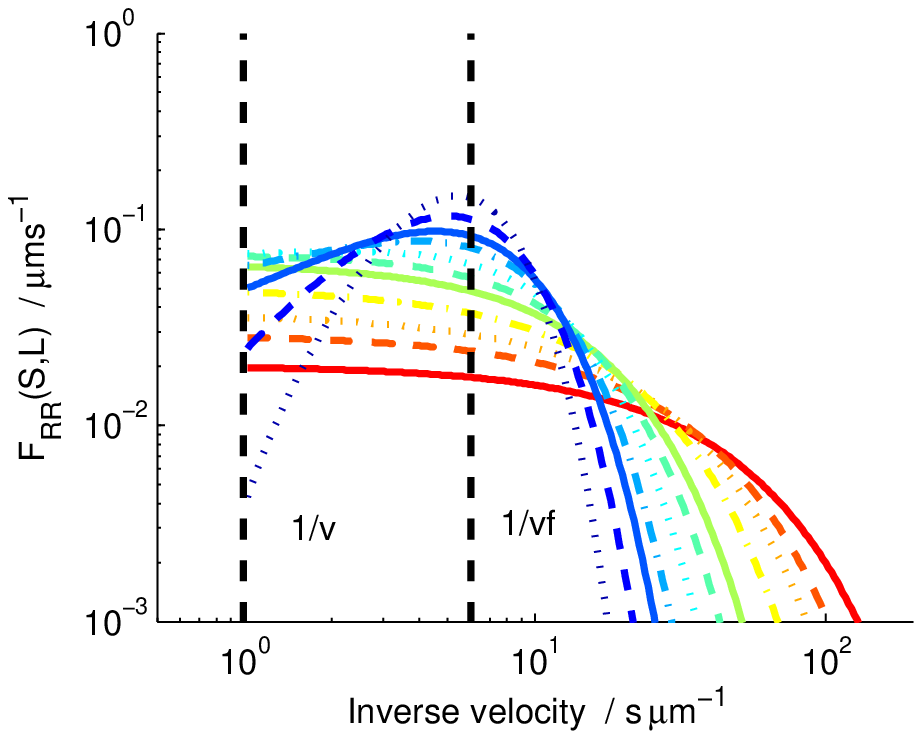}}}
\subfigure[]{\resizebox{8cm}{!}{\includegraphics{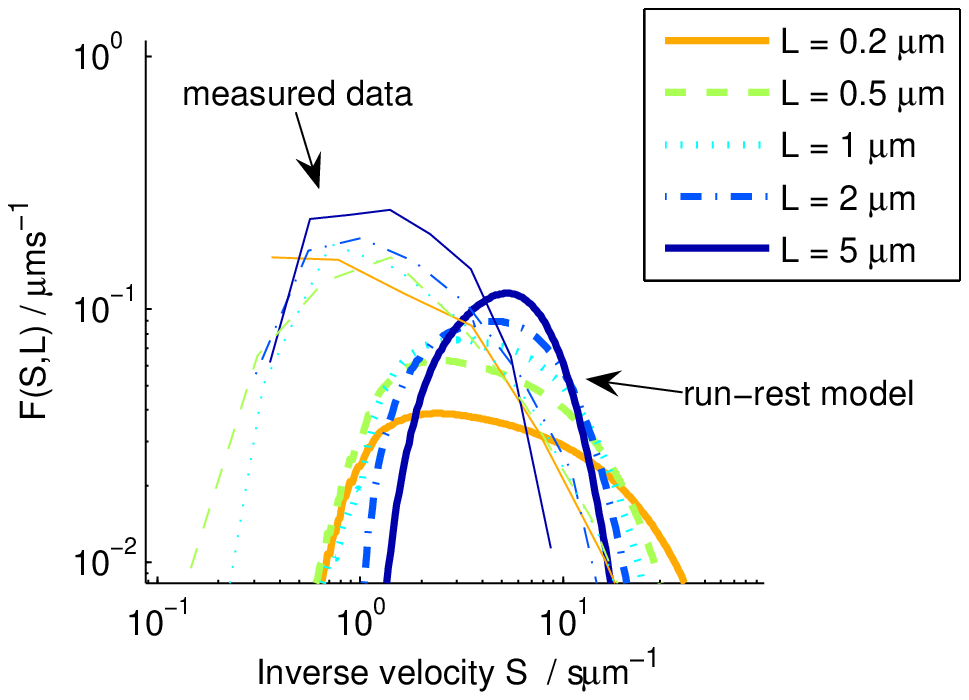}}}
\caption{The first passage probability for the ``run-rest'' model, taking parameters $v$, $a$ and $b$ from the segmentation analysis of the control data. (a) Eq.~\ref{eq:rr} is evaluated with $v=1.0~\mu$m~s$^{-1}$, $a^{-1}=0.76$~s, and $b^{-1}=3.8$~s: $F_{\textrm{(RR)}}$ switches from a delta function (-\,-\,-) at $1/v=1~$s~$\mu$m$^{-1}$ when $L$ is small, towards a delta function at $1/(vf)=6.0~$s~$\mu$m$^{-1}$ when $L$ is large. Colours correspond to the legend of Fig.~\ref{fpp_all}. (b) Eq.~\ref{eq:rr} is convolved with the distribution of run speeds in Fig.~\ref{segmentation}(a) above, again taking $a^{-1}=0.76$~s, and $b^{-1}=3.8$~s. The predicted $F(S,L$ agrees qualitatively with the measurement, showing a narrowing of the distribution with increasing $L$.}
\label{runrest}
\end{figure}

We compare the run-rest model with the observed data by taking the relevant parameters of Eq.~\ref{eq:rr} from a segmentation analysis of the observed data, as described above. Fig.~\ref{segmentation} shows statistics of the lengths, times and speeds of runs and rests in the control cells, from which we measure $v=1.0~\mu$m~s$^{-1}$, the mean speed of the segmented runs, $a^{-1}=0.76$~s, the mean lifetime of the segmented runs, and $b^{-1}=3.8$~s, the mean lifetime of the segmented rests. Taking these values of $v$, $a$ and $b$, $F_{\textrm{(RR)}}(S,L)$ is plotted in Fig.~\ref{runrest}(a). The predicted FPP does indeed switch from a delta peak at $S=v^{-1}$ to a delta peak at $S=(vf)^{-1}=6.0$~s~$\mu$m$^{-1}$. Having seen that Fig.~\ref{segmentation}(a) shows a distribution of run speeds $P(v)$, we then convolve $F_{\textrm{(RR)}}$ with $P(v)$ to compare directly our run-rest model with $F_r(S,L)$ from the measured control data (Fig.~\ref{runrest}(b)). The run-rest model agrees qualitatively with the measured data, showing a similar narrowing with increasing $L$. However, there is a quantitative discrepancy: the predicted inverse velocities are about a factor of 5 lower than the measured values.

Since calculation of the first passage probability from measured particle tracks is a straightforward procedure, which employs no assumptions about the particle motion, nor arbitrary parameters, this comparison between the first passage probability analysis and the run-rest model shows that the latter does not accurately reflect the motion of the endocytic vesicles. Nevertheless, we do not wish to devalue the segmentation of tracks into runs and rests, because, as others have found, it still a useful analysis of particle motion, as it is conceptually simple and motivated by direct observation. However, we discuss below some possible reasons why the run-rest model is inadequate for predicting the first passage probability distribution of endocytic vesicle motion.

\section{Discussion}
We believe that the concept of first passage probability, of particles to be transported a certain distance, is a natural way of analysing the trajectories of trafficked intracellular particles, especially when our theoretical knowledge of their motion is incomplete. The calculation of FPP reduces the observed tracks without employing assumptions about the nature of each particle's motion, nor arbitrary parameters and thresholds. It also leads to a useful graphical interpretation of measured data: the distribution that results are seen to be qualitatively different for conceptually different types of motion, such as random walks, directed motion, and the run-rest model. In comparing measured data, the FPP has allowed us to visualise differences in motion, in cells treated with nocodazole compared to control cells (Fig.~\ref{compare_fpp}, Table \ref{ref_fpp}). By comparing the FPP, we have seen that each cell treatment has markedly reduced the probability of vesicles being trafficked at higher speeds and larger distances. The FPP has also allowed us to visualise the reproducibility of motion between different cells (Fig.~\ref{variability}). 

The FPP can be calculated as a function of different variables. From the point of view of the theory, the most natural is time, because for each position $\mathbf{R}_n(T)$ along a track $n$, we measure the time that the particle takes to make its first passage of length $L$ from that position. However, from the point of view of measuring trafficking, in which particle trajectories bear some resemblance to ballistic, unidirectional motion, it is natural to plot the FPP as a function of inverse velocity $S=t/L$, or velocity $L/t$, since either allows us to instantly recognise steady unidirectional motion. We choose $F(S,L)$ for two reasons: inverse velocity is closer to the calculation because it only involves scaling the times by $L$; and a function of inverse velocity shows faster moving particles more prominently than a function of velocity, because of the implicit transformation between the two (Eq.~\ref{implicitTransformation}).

Three very simple models of particle motion have been compared with the measured FPP. The two extremes of steady directional motion and diffusion led to predictions that were qualitatively different from the measured FPP, so a quantitative comparison was unnecessary. The run-rest model, of constant run velocity $v$, and constant ``on'' and ``off'' rates, $b$ and $a$, predicted a FPP that was qualitatively similar to the measured data, when we selected $a$ and $b$ from a segmentation analysis of the measured data, and allowed a convolution of $v$ over the distribution of run speeds from the segmentation analysis. However, the predicted values of $S$ were approximately a factor of 5 from the measured distribution. There are, of course, many sources of discrepancy between the run-rest model and the experiment; some were mentioned above: abrupt changes of direction, reversals of motion along a linear track, curvature of the path and slower erratic motion. It is outside the scope of this study to model these effects properly: obviously, to model the observed vesicle motion properly would require a much deeper understanding of the governing mechanisms.

However, we suspect that there may be simple interpretations of the coarse discrepancy that we have found. For instance, we believe that is wrong to assume there are only two rates $a$ and $b$ that govern switching from run to rest and vice versa. For example, a run could be interrupted by various different physical processes such as detachment of the motor, obstruction of the motor or load, reaching the end of the microtubule, binding of the motor to a regulatory protein, etc. We see a few fast-moving tracks that are interrupted by frequent short rests, as well as particles which spend most of their time sessile, but make occasional jumps. The fact that there must be more than one type of rest is reflected in the non-exponential rest time distribution of Fig.~\ref{segmentation}(d). 

We must also be aware that it is difficult to assess the reliability of the segmentation analysis, especially where the particle tracks do not follow straight lines, and are complicated by several types of motion as well as measurement error. For example, our segmentation analysis ignores runs that are short in time or length, or slow in speed: these runs would be counted instead as part of a rest. Rests that are short in time are also ignored because of the thresholds. Likewise, smoothing of particle tracks ``irons out'' short runs and short rests, making runs and rests appear longer in time, and runs appear slower in speed. The process of thresholding and smoothing can therefore artificially change our predictions of $a$, $b$ and $P(v)$ to a significant degree that could easily be missed. On the other hand, the FPP analysis is transparent: no smoothing or thresholding are involved, and the measured tracks yield a first passage probability distribution in a single mathematical step.

\section{Conclusion}
The first passage probability distribution is a robust measure of intracellular particle tracks that allows us to reduce the measured data into a form in which qualitative and quantitative differences in motion can be easily visualised and compared. The FPP is transparent in that it does not rely on arbitrary parameters or thresholds, nor any implied assumptions about the mechanism of a particle's motion, unlike the alternative segmentation analyses. Therefore it is an attractive way of analysing intracellular particle trafficking which is neither completely random nor completely deterministic. The FPP gives us an illuminating way of comparing theories of trafficking with experiment.


\bibliography{FirstPassageProb-article}
\end{document}